\documentclass[preprint,showpacs,preprintnumbers,amsmath,amssymb]{revtex4}
\usepackage{graphicx}
\usepackage{dcolumn}
\usepackage{bm}
\usepackage{amssymb}
\usepackage{amsfonts}
\usepackage{amsmath}
\usepackage{textcomp}
\usepackage{color}
\UseRawInputEncoding

\begin{document}

\title{Screening effect in Spin-Hall Devices}
\author{M. Creff}
\affiliation{LSI, \'Ecole Polytechnique, CEA/DRF/IRAMIS, CNRS, Institut Polytechnique de Paris, 91120 Palaiseau, France}
\author{E. Olive}
\affiliation{GREMAN UMR-CNRS 7347, UniversitŽ de Tours, INSA Centre Val de Loire, Parc de Grandmont, 37200 Tours, France}
 \author{J.-E. Wegrowe} \email{jean-eric.wegrowe@polytechnique.edu}
\affiliation{LSI, \'Ecole Polytechnique, CEA/DRF/IRAMIS, CNRS, Institut Polytechnique de Paris, 91120 Palaiseau, France}
\date{\today}

\date{\today}

\begin{abstract}
The stationary state of the spin-Hall bar is studied in the framework of a variational approach that includes non-equilibrium screening effects. The minimization of the power dissipated in the system is performed with taking into account the spin-flip relaxation and the global constrains due to the electric generator and global charge conservation. The calculation is performed in both approximations of negligible spin-flip scattering and strong spin-flip scattering. In both cases, the expressions of the spin-accumulation and the longitudinal and transverse pure spin-currents are derived analytically. Due to the small value of the Debye-Fermi screening length, the spin-accumulation is shown to be linear in $y$ (across the device), linear in the electric field imposed by the generator, and inversely proportional to the temperature for non-degenerate conductors.
\end{abstract}


\maketitle

\section{Introduction}
Spin-accumulation can be produced at the edges of a conducting bar at zero external magnetic field, by injecting an electric current into a non-magnetic material with high spin-orbit coupling \cite{Awschalom,Jungwirth,Valenzuela,Otani,Gambardella,Bottegoni}. This effect is called Spin Hall Effect (SHE). It has been predicted some decades ago \cite{Dyakonov}, and described in the framework of various theoretical models \cite{Dyakonov2,Hirsch,Zhang, Tse,Maekawa, Hoffmann,Saslow,SHE,Sinova}. However, a description that takes into account the {\it non-equilibrium nature of the electric screening} occurring in the SHE is still an open problem. One of the main difficulty is the same as for the classical Hall-effect \cite{boundary,Trefenthen,Heremans,Geometry,Connection,Perrott,Nanowire,Calcul,Solin,SolinJAP}: it is due to the fact that the values of the charge accumulation at the edges of the Hall bar are not directly imposed by the external constraints. Instead, for the stationary state, the accumulation of electric charges and spins at the edges is generated by the system itself, in reaction to the action of the internal magnetic field (the spin-orbit effective magnetic field),  together with to stationary current injected from the electric generator, and for a given geometry. As a consequence, the determination of a unique solution to the drift-diffusion equations at stationary state based on the {\it local boundary conditions} is problematic \cite{Tse}.\\

As shown in a series of publications that led to the present work\cite{EPL1,EPL2,Benda,JAP1,JAP2,JAP3} - it is possible to take into account the non-equilibrium nature of the electric screening in the case of a ideal Hall bar on the basis of the least dissipation principle \cite{Onsager_Diss,Bruers,MinDiss}. The stationary state can then be deduced from the minimization of the power under the global constraints, instead of solving independently the drift-diffusion equations. This approach has been applied successfully in the case of the classical Hall effect without spin \cite{JAP1,JAP2,JAP3}. It has been shown that the charges accumulated at the edges are not static but generate a {\it non-uniform longitudinal current} $\delta J_x(y)$, which has not been derived before. Physically, this surface current is responsible for the well-known robustness of the Hall voltage, which can be measured while using a good or a bad voltmeter (i.e. with huge variations of the electric leakage at the edges \cite{Hall}), because the electric charges accumulated are renewed permanently despite the zero transverse current. Furthermore, if a secondary passive circuit is contacted the two edges of the Hall bar, the variational method shows that the electric current injected is mainly carried by the longitudinal surface currents $\delta J_x(y)$, instead of the transverse current $J_y$  \cite{JAP3}.\\

The model developed in the present work is the application of the variational method to the SHE. The model is based on two spin-channel description of the electric currents, that is well-known in the studies of spin transport in the context of giant magnetoresistance \cite{Johnson,Wyder,Fert,PRB2000,Schmidt,Jedema,JPhys17}. The spin-Hall bar is then equivalent to a superimposition of two subsystems composed of two classical Hall bars, activated by an effective spin-orbit magnetic field. The sub-systems are the two spin-dependent electronic carriers, and the effective magnetic field is acting with identical amplitude and opposite sign on each spin-channel. The consequence of these symmetries on the electric transport is that the electric charges accumulated at the edges have an opposite sign for each channel, so that the sum over the two channel is zero (if the symmetry is strictly respected), and the difference gives the spin-accumulation observed. The application of the variational method presented below shows that the same symmetry also applies for the out-of-equilibrium spin currents, as sketched in Figure 1 and Figure 2. The transverse and the non-homogeneous longitudinal currents are both pure spin-currents.\\

The paper is organized in three sections: after the presentation of the model and formalism (Section II below), the derivation of the spin-accumulation and currents is performed in the case of negligible spin-flip scattering in section III. The Joule functional is minimized under the  global constrains, and the stationarity condition for the electric current $\vec J$ is obtained. The accumulation of electric charges and spin are then derived by integration of the Maxwell-Gauss equation. In section IV, the same derivation is performed in the case of strong spin-flip scattering, by adding the spin-flip dissipation into the Joule functional. In order to deal with analytical expressions, the spin-flip scattering length $l_{sf}$ is assumed to be small with respect to the width of the Hall-bar $\ell$. It is shown that there is a solution of continuity between the two limits ($\ell \ll l_{sf}$ and $\ell \gg l_{sf}$). From a more quantitative viewpoint, the small value of the Debye length scale $\lambda_D$ leads to the following surprising result: the spin-flip scattering does not modify significantly the profile of the spin-accumulation $n^{Sp}(y)$, nor that of the longitudinal spin-current $J^{Sp}_x(y)$. Three main properties of the spin-accumulation should thus be expected: (a) the linearity with $y$, (b) the linearity with the applied voltage $\Delta V$, and (c) the $1/T$ dependence to the temperature for non-degenerate conductors (as an expression of the paramagnetic behavior of the spin-accumulation). The results are summarized in the conclusion (section V).

\begin{figure} [ht]
   \begin{center}
   \includegraphics[height=7cm]{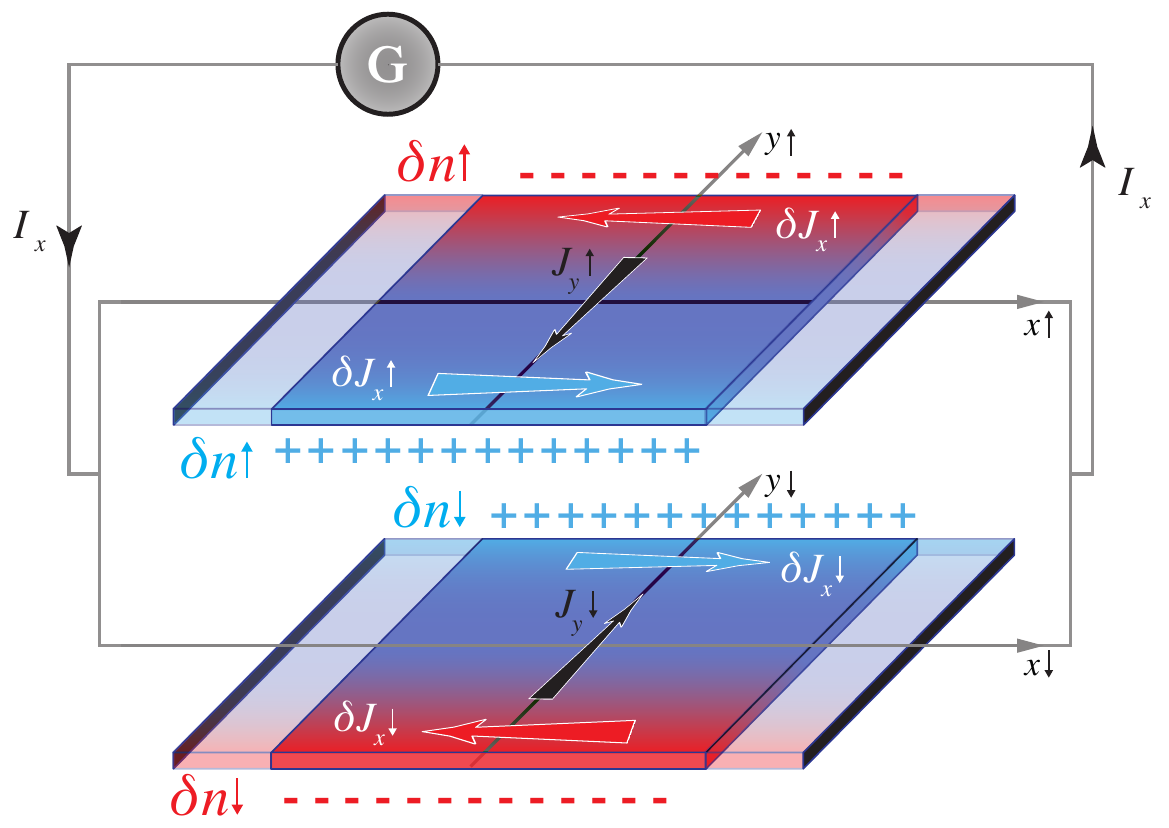}
   \end{center}
   \caption[example] 
   { \label{fig:Fig.2} 
Schematic representation of the two channel model (corresponding to spin $\uparrow$ and spin $\downarrow$) for a spin-Hall-bar. The device is contacted to an electric generator that imposes a longitudinal stationary current $J_x^0$ (not shown). The charge accumulations $\pm \delta n_\updownarrow$ and the inhomogeneous part of the longitudinal currents $\delta J_{x \updownarrow}(y) = J_{x \updownarrow}(y) - J_x^0$ are represented at both edges for each channel, together with the transverse current $J_{y \updownarrow}$ flowing from one edge to the other.}
   \end{figure}

\begin{figure} [ht]
   \begin{center}
    \includegraphics[height=5cm]{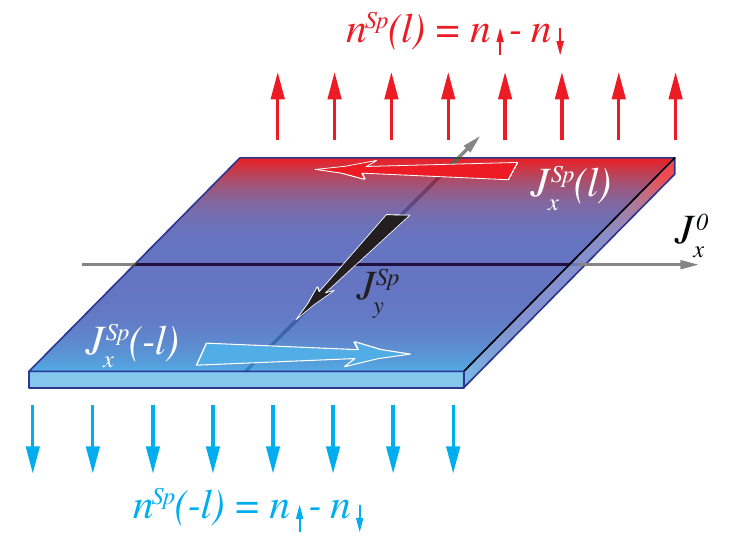}
   \end{center}
   \caption[example] 
   { \label{fig:Fig.1} 
   Superimposition of the two channels represented for symmetric device and environment ($C_E = 0$ in the text). The electric generator and the injected current $J_x^0$ are not shown for clarity. The system is characterized by the absence of total charge accumulation,  the spin-accumulation $n^{Sp}(y) = \delta n_{\uparrow }(y) - \delta n_{\downarrow }(y)$ with opposite directions at the two edges, the inhomogeneous part of the longitudinal spin-currents $J^{Sp}_x(y) = J_{x \uparrow }(y) - J_{x \downarrow }(y)$, and the transverse spin-current $J^{Sp}_y = J_{y \uparrow }(y) - J_{y \downarrow }(y)$.  }
   \end{figure} 

\section{The model}

The system is a long bar of width $\ell$ and thickness $d$, composed of a thin layer of a non-magnetic material with strong spin-orbit coupling contacted to an electric generator. We assume that the bar is invariant by translation along its longitudinal axis $x$ (i.e. we do not take into account the region close to the generator, nor the perturbation caused by possible lateral contacts), that the lateral edges are symmetric, that the device is planar (no charge transport in the $z$ direction), and that the polarization axis of the spins is oriented along the $z$ direction. In the framework of the two channel model, the charge carriers are separated into two populations, that are the charge carriers of spin up $\uparrow$ and the charge carriers of spin down $\downarrow$, with the respective charge densities $n_\uparrow$ and $n_\downarrow$.  

The conductor is characterized by a density of charge carriers $n^Q \equiv n_\uparrow + n_\downarrow = 2 n_0 + \delta n^Q$, where $n_0$ is the density defined by electroneutrality at equilibrium (i.e. without current injected). Accordingly $n_0$ is uniform and does not depend on the spin-channel. On the other hand $\delta n^Q(y)$ is the charge accumulation, which is not a function of $x$ due to the invariance by translation, and is the sum of the charge accumulation for each spin-channel: $\delta n^Q(y) = \delta n_\uparrow(y) + \delta n_\downarrow(y) $.
The Coulomb interaction is described by the electric potential $V$, that follows the Poisson law $\nabla^2 V = - q \delta n^Q /\epsilon$, where  $q$ is the charge of the carriers, $\epsilon$ is the permittivity of the material, and $\vec \nabla = \{\partial_x, \partial_y\}$ is the gradient operator in 2D. 

 The energy of the system is then defined by the two chemical potentials $\mu_{\uparrow}$ and $\mu_{\downarrow}$ such that:
\begin{equation}
\mu_{\updownarrow} = \frac{kT}{q}\,  \ln{\left (\frac{n_{\updownarrow}}{n_0} \right )} + V + \mu_{\updownarrow}^{ch}
\label{ChemPot}
\end{equation}
where $k$ is the Boltzmann constant, $T$ is equal to the temperature of the material in the case of non-degenerate semiconductors, or $T$ is equal to the Fermi temperature $T\equiv T_F$ in the case of degenerate semiconductors and metals \cite{RqueT}. The first term in the right hand side accounts for the diffusion of the carriers (this term is justified in the framework of the local equilibrium approximation \cite{Rubi,JPhys17}).  The second term describes the electric potential $V$. The third term $\mu_{\updownarrow}^{ch}$ is the ÔÕpure chemical potentialÕÕ that accounts for the spin-flip scattering, and is the main parameter for the description of giant magnetoresistance effects \cite{Johnson,Wyder, Fert, Jedema, Schmidt,PRB2000}. In the following, we assume that $\mu_{\updownarrow}^{ch}$ is uniform in order to treat uniquely the spin-accumulation due to the SHE. The electric field reads $\vec {\mathcal E} = - \vec \nabla V =  \mathcal E_x^0 \, \vec e_x + \mathcal E_y \, \vec e_y $, where the $x$-component $\mathcal E_x^0$ is constant (due to the invariance along $x$), so that the Poisson law is reduced to $\partial_y \mathcal E_y = q \delta n^Q/\epsilon$. Note that the electric field does not depend on the spin-channel, so that the electric potential  $V$ couples the two channels in Eq.(\ref{ChemPot}).
We can re-write the Poisson law with the help of the chemical potential Eq.(\ref{ChemPot}) as follow:
\begin{equation}
\nabla^2 \mu_{\updownarrow} - \lambda_D^2 \, \frac{q n_0}{\epsilon}\, \nabla^2  \ln{\left (\frac{n_{\updownarrow}}{n_0} \right )} + \frac{q \delta n^Q}{ \epsilon} = 0
\label{PoissonChem}
\end{equation}
where $ \lambda_D = \sqrt{\frac{kT \epsilon}{q^2 n_0}}$ is the well-known Debye-Fermi screening length.

On the other hand, the transport equations for the charge carriers are given by the Ohm's law for each channels:
\begin{equation}
\vec{J}_{\updownarrow} = -\hat \sigma_{\updownarrow} \, \vec{\nabla} \mu_{\updownarrow} = - q n_{\updownarrow} \hat \eta_{\updownarrow} \, \vec{\nabla} \mu_{\updownarrow},
\label{Ohm}
\end{equation}
where $\hat \sigma_{\updownarrow} \equiv q n_{\updownarrow} \hat \eta_{\updownarrow}$ is the spin-dependent conductivity tensor, and $\eta_{\updownarrow}$ is by definition the electric mobility tensor. 

As a consequence of both the Onsager reciprocity relations \cite{Onsager} and the property of the spin-orbit effective fields, the mobility tensor takes the following form in the orthonormal basis $\{\vec e_{x_{\uparrow}},\vec e_{y_{\uparrow}}, \vec e_{x_{\downarrow}},\vec e_{y_{\downarrow}}\}$ (see Fig.1):
\begin{equation}
  \hat{\eta}_{\updownarrow}
  = \eta
   \begin{pmatrix}
    1    & \theta_{so} & 0 & 0\\
    -\theta_{so} & 1 & 0 & 0 \\ 
   0 & 0 & 1  & - \theta_{so} \\
    0 & 0 & \theta_{so} & 1\\ 
  \end{pmatrix}
      \label{Mobility}
  \end{equation}
where $\eta$ is the mobility of the material and $\theta_{so}$ is the spin-Hall angle.
 the transport equation reads:
\begin{equation}
\vec{J}_{\updownarrow}  = - q n_{\updownarrow}  \eta \left( \vec{\nabla} \mu_{\updownarrow}  \mp \theta_{so} \, \vec{e_z} \times \vec{\nabla} \mu_{\updownarrow}  \right),
\label{Ohm2}
\end{equation}
This equation is equivalent to the Dyakonov-Perel equation of the spin-Hall effect \cite{JPhys17}. It is convenient to rewrite Eq.(\ref{Ohm2}) in the following forms:
\begin{align}
  -q n_{\updownarrow}  \eta (1 + \theta_{so}^2) \partial_x \mu_{\updownarrow}  & = J_{x \updownarrow}  \mp \theta_{so} J_{y \updownarrow}  
  \label{relations-dxmu} \\
  -q n_{\updownarrow}  \eta (1  + \theta_{so}^2) \partial_y \mu_{\updownarrow}  & = J_{y \updownarrow}  \pm \theta_{so} J_{x \updownarrow}
  \label{relations-dymu} \\
\|\vec J_{\updownarrow} \|^2  \equiv J_{x \updownarrow} ^2 + J_{y \updownarrow} ^2 &  = (qn_{\updownarrow}  \eta)^2 \, (1 + \theta_{so}^2) \,  \|\vec \nabla \mu_{\updownarrow}  \|^2.
\label{OhmHall3}
\end{align}
where the symbols $\mp$ and $\pm$ refer to the spin-channel $\uparrow$ for the upper sign and spin-channel $\downarrow$ for the lower sign.
 The power dissipated by the charge carriers in each spin-channel is the Joule power : 
\begin{equation}
  P_{J \updownarrow}  = S_\text{lat} \int_{-\ell}^{\ell}  q n_{\updownarrow}  \eta \|\vec{\nabla} \mu_{\updownarrow}  \|^2 d y
       = \frac{S_{lat}}{q n_0 \eta (1+\theta_{so}^2)}\int_{-\ell}^{\ell} \frac{n_0}{n_{\updownarrow} } \|\vec{J}_{\updownarrow} \|^2 d y.
       \label{JoulePower}
\end{equation}
where we have introduced the lateral surface $S_{lat} = L_x \, d$, where $L_x$ is the length of the Hall bar along the $x$ direction.\\

On the other hand, the power dissipated by the spin-flip scattering (i.e. the transition of an electric charge from one spin-channel to the other) per unit of volume reads $\mathcal L \, \Delta \mu^2$, where $\Delta \mu = \mu_{\uparrow} - \mu_{\downarrow}$ and $\mathcal L$ is the Onsager transport coefficient describing the spin-flip relaxation rate in the framework of the two channel model\cite{Wyder,Fert,PRB2000,JPhys17,EPL2}. Note that the relaxation rate $\dot \psi = \mathcal L \Delta \mu$ is by definition the source term in the continuity equation for the spin-dependent electric charges (in the absence of charge accumulation): $q\partial n_{\updownarrow}/ \partial t = - \vec \nabla .\vec J_{\updownarrow} \mp \dot \psi$. Accordingly, the coefficient $\mathcal L$ is related to the measured {\it spin-flip scattering length} $l_{sf}$  by the relation \cite{JPhys17}
\begin{equation}
\mathcal L = \frac{\sigma_0}{2 l_{sf}^2},
\label{SF_length}
\end{equation}
where $\sigma_0$ is the conductivity of the material.
In the framework of the variational approach, and ignoring temperature gradient, the stationary state is defined by the principle of {\it least power dissipation }\cite{Bruers,MinDiss}. In other terms, the distribution of charge densities and currents is determined by the state of minimum power production, taking into account the global constraints applied to the system. These constraints are due to the presence of the electric generator, the electrostatic boundary conditions, and the symmetries of the device \cite{EPL1,EPL2,JAP1,JAP2,JAP3}. Due to the {\it electroneutrality} and the fact that the device is symmetric, the integral of the total charge accumulated in the two channels cancels out $\int_{- \ell}^{\ell} ( \delta n_{\uparrow} + \delta n_{\downarrow} )dy = 0$. Or in other terms: 
 \begin{equation}
\frac{1}{2 \ell} \, \int_{- \ell}^{\ell} \left ( n_{\uparrow}(y)+  n_{\downarrow}(y) \right ) dy  = 2  n_0
\label{ConstrQ}
 \end{equation}
On the other hand, the generator injects a constant current through the device (of constant lateral surface $S_{lat}$), si that the integrated current is constant:
  \begin{equation}
 \frac{1}{ 2  \ell} \,  \int_{- \ell}^{\ell} \left ( J_{x \uparrow}(y) + J_{x \downarrow}(y) \right ) dy  = 2  J_x^0
   \label{ConstrJ}
 \end{equation}


\section{Spin Hall Effect with negligible spin-flip scattering}

The approximation of negligible spin-flip scattering corresponds to the case $l_{sf} \gg \ell$ where $\ell$  is the width of the spin-Hall bar. Let use define for convenience the reduced power:
$\tilde{P}_{J \updownarrow} = \frac{q \eta (1+\theta_{so}^2)}{S_\text{lat}} \, P_{J \updownarrow}   = \int_{- \ell}^{\ell} \frac{J_{x \updownarrow}^2 + J_{y \updownarrow} ^2}{n_{\updownarrow}} dy$.
We shall use the Lagrange multipliers $\lambda_J$ and $\lambda_n$ in order to take into account the constraints respectively Eq.(\ref{ConstrQ}) and Eq.(\ref{ConstrJ}) in the minimization of the functional $\mathcal F$:
 \begin{equation}
{\mathcal{F}}_{J \updownarrow} [J_{x \updownarrow},J_{y \updownarrow},n_{\updownarrow}] =   \int_{- \ell}^{\ell} \left ( \frac{J_{x \updownarrow}^2 + J_{y \updownarrow}^2}{n}  - \lambda_J \, \left ( J_{x \uparrow} + J_{x \downarrow} \right ) - \lambda_n \, \left ( n_{\uparrow} + n_{\downarrow}  \right ) \right ) dy 
 \end{equation}
The minimum of the reduced Joule power ${\mathcal{F}}_{J} = \tilde {\mathcal{P}}_{J \uparrow} + \tilde {\mathcal{P}}_{J \downarrow}$ corresponds to :
 \begin{equation}
  \frac{\delta \mathcal{F}_{J \updownarrow}}{\delta J_{x \updownarrow}} = 0  \ \Longleftrightarrow \ 2 J_{x \updownarrow} = n_{\updownarrow} \lambda_J,
 \label{CondJx00}
\end{equation}
 \begin{equation}
  \frac{\delta \mathcal{F}_{J \updownarrow}}{\delta J_{y \updownarrow}} = 0 \ \Longleftrightarrow \ J_{y \updownarrow} = 0,
\end{equation}
 \begin{equation}
  \frac{\delta \mathcal{F}_{J \updownarrow}}{\delta (n_{\updownarrow})} = 0 \ \Longleftrightarrow \ J_{x \updownarrow}^2 + J_{y \updownarrow}^2 = - \lambda_n n_{\updownarrow}^2,
 \label{Condn}
\end{equation}
Using Eqs.(\ref{ConstrQ}), Eqs.(\ref{ConstrJ}),  and Eq.(\ref{CondJx00}) leads to $\lambda_J = \frac{2 J_x^0}{n_0}$ so that $J_{x \updownarrow}  = \frac{n_{\updownarrow}}{n_0} J_x^0 $ (and from Eq.(\ref{Condn}) we have furthermore $\lambda_n = - (J_x^0 / n_0)^2$). The stationarity conditions for the currents are then defined by the two relations:
\begin{equation}
  J_{x \updownarrow}(y) = J_x^0\frac{n_{\updownarrow}(y)}{n_0} \quad \text{and} \quad J_{y \updownarrow} = 0.
  \label{min}
\end{equation}
 Like for the usual Hall effect, {\it there is no transverse current} in the device for $\l_{sf} \gg \ell$. Let us define the inhomogeneous current $\delta J_{x \updownarrow} =J_{x \updownarrow} -  J_x^0$ produced by the spin-orbit field. This longitudinal spin-current is a {\it pure spin current}, in the sense that the unhomogeneous current of electric charge  $\delta J_x^Q$ is zero and the current of spins $J_x^{Sp}$ is proportional to the spin accumulation $n^{Sp}(y)$:
\begin{equation}
 \delta J_x^Q = \delta J_{x \uparrow} + \delta J_{x \downarrow} = 0 \qquad \text{and}  \qquad J_x^{Sp} = \delta J_{x \uparrow} - \delta J_{x \downarrow} = J_x^0 \frac{n^{Sp}}{n_0} .
 \label{PureSpinCurrent_x}
 \end{equation}
 
The explicit expression of $ J_{x \updownarrow}(y)$ is obtained if we know the expression of the densities $n_{\updownarrow}(y)$. This is the aim of the following paragraph.

Inserting the stationarity conditions Eq.(\ref{min}) into Eq.(\ref{relations-dxmu}) and Eq.(\ref{relations-dymu}) we deduce $\partial_x \mu_{\updownarrow}  =  \partial_x \mu =   \frac{-J_x^0}{q n_0 \eta (1+\theta_{so}^2)}$ and $\partial_y \mu_{\updownarrow} =  \mp \frac{\theta_{so} J_x^0}{q n_0 \eta (1+\theta_{so}^2)}$. These two terms are constant so that $\nabla^2 \mu_{\updownarrow} = 0$ and the chemical potentials $\mu_{\updownarrow}$ are harmonic functions. Equation (\ref{PoissonChem}) reduces to:
\begin{equation}
 \lambda_D^2 \partial_y^2 \ln \left(1 + \frac{\delta n_{\updownarrow} }{n_0} \right) = \frac{\delta n^Q}{n_0}.
  \label{poisson-final}
\end{equation}
In order to evaluate the global boundary conditions for the electric field we have to integrate the Maxwell equation $\vec{\nabla} \cdot \vec{\mathcal E} = \partial_y \mathcal E_{y}  = \frac{q}{\varepsilon} \delta n^Q$ at a given point $y_0$ inside the material :
\begin{equation}
  \mathcal E_{y}(y_0) = - \partial_y V(y_0) = \frac{E^{\infty}}{2} - \frac{q}{2 \varepsilon} \int_{-\ell}^{\ell} \delta n^Q(y) sgn(y-y_0) dy ,
  \label{ElecField}
\end{equation}
where $E^{\infty}$ accounts for possible electric charges in the environment of the spin-Hall bar (this is the case if a conducting or insulating layer is deposited on one side of the spin-Hall bar). Inserting the chemical potential Eq.(\ref{ChemPot}), the transport equation Eq.(\ref{relations-dymu}), and the stationary conditions Eq.(\ref{min}), yields:
\begin{equation}
 \pm \lambda_{SH}^{0} + \lambda_D^2 \partial_y \ln \left (\frac{n_{\updownarrow}}{n_0} \right )(y_0) - \frac{C_E}{2}  + \frac{1}{2 n_0 } \, \int_{-\ell}^{\ell} \delta n^Q(y) sgn(y-y_0)d y = 0,
  \label{condition}
\end{equation}
where the asymmetry of the electric environment is described by the characteristic length $C_E = \frac{\varepsilon E^{\infty}}{q n_{0}}$. We have also introduced {\it the spin-Hall characteristic length}: 
\begin{equation}
\lambda_{SH}^{0} = \frac{\varepsilon }{q^2 n_0^2 \eta} \, \frac{ \theta_{so} }{1 + \theta_{so}^2} \, J_x^0 = \frac{\lambda_D^2 }{kT } \left ( \frac{\theta_{so}}{1 + \theta_{so}^2} \right ) \frac{J_x^0}{n_0 \eta}.
\label{lambdaSH}
\end{equation} 

At the first order in $\delta n_{\updownarrow}/n_0$, the difference of both spin-channels in Eqs.(\ref{condition}) gives the spin-accumulation $n^{Sp} = \delta n_{\uparrow} - \delta n_{\downarrow}$, linear in $y$:
\begin{equation}
  \frac{n^{Sp}(y)}{n_0} =   - 2 \frac{ \lambda_{SH}^{0}}{ \lambda_D^2} \, y =  - 2  \left ( \frac{\theta_{so}}{1 + \theta_{so}^2} \right ) \, \left ( \frac{q \Delta V}{kT} \right ) \frac{y}{L},
  \label{spin-accum}
\end{equation}
where we have introduced on the right hand side the longitudinal voltage $\Delta V$ in order to compare the energy imposed by the generator $q \Delta V$ with the thermal energy $kT$, as illustrated in Figure 3b. The voltage $\Delta V$ is measured along the axis $x$ over the distance $L$, and is given by the relation $ \Delta V = \mathcal E_x^0 L  = L \,J_x^0/\sigma_0 $. 
 
 Note that, according to Eq.(\ref{spin-accum}), the spin-current $J^{Sp}_{x}$ (Eq.(\ref{PureSpinCurrent_x})) is proportional to the square of the injected current density $\left ( J_x^0 \right)^2$. The Joule heating is thus expected to be a power four $\left ( J_x^0 \right)^4$, which is not without consequence for the temperature distribution inside the device \cite{Beach}.\\
In order to summarize, we see that three main properties of the spin-accumulation are derived in the limit $l_{sf} \gg \ell$: (a) the linearity with $y$, (b) linearity with the applied voltage $\Delta V$, and (c) proportionality with $1/T$ for non-degenerate conductors. 
From a more quantitative viewpoint, assuming a small sample in which $L$ and $\ell$ are of the same order, the spin-accumulation at the border is of the order of $n^{Sp}(y = \ell)/n_0 \approx \theta_{so} \, q \Delta V / kT$, where $\theta_{so}$ is of the order of $10^{-3} -10^{-4}$ and the maximum of $q \Delta V$ is a fraction of $eV$.  Accordingly, the ratio $n^{Sp}(y = \ell)/n_0$ can reach  $0.1$, e.g. at low temperature (as shown in  Figure 3b).\\

Furthermore, if both the device and its environment are {\bf  symmetric} we have $C_E = 0$. The sum of the two Eqs.(\ref{condition}), at the first order in $\delta n_{\updownarrow}/n_0$, gives 
\begin{equation}
  \frac{\delta n^{Q}(y)}{n_0} =  0,
  \label{charge-accum0}
\end{equation}
and the relation $\delta n_{\uparrow} = - \delta n_{\downarrow}$ is verified. This corresponds to symmetric SHE devices, as sketched in Fig.1 and Fig.2. \\
 
However, if the device or its environment are {\bf not symmetric} (typically if a dielectric is placed on one side of the device),  $C_E \ne 0$ and there is an asymmetry of the charge accumulation and spin-polarization between the two edges. The following charge accumulation appears :
\begin{equation}
  \frac{\delta n^Q (y)}{n_0} =    \frac{C_E}{ \lambda_D \sqrt{2}} \frac{sh \left ( \frac{\sqrt{2}}{\lambda_D} y \right )}{ch \left ( \frac{\sqrt{2}  }{  \lambda_D} \ell\right )},
   \label{charge-accum}
\end{equation}
and residual spin-Hall voltage is generated. This result seems to be in agreement with that derived in the reference \cite{DiVentra}.

It is worth noting that if a ferromagnetic layer is present on one edge of the spin-Hall bar (e.g. a ferromagnetic 3d metal, or a magnetic oxide like YIG) or a spin-injection mechanism is present also on one edge of the spin-Hall bar, the charge accumulation $\delta n^{Q}$ will depend on the magnetization state $\vec M$ of the ferromagnet, or on the state of the spin-injector (also denoted by $\vec M$ for convenience). The asymmetry of $\delta n_{\updownarrow}$ will then lead to a spin-dependent voltage where the parameter $C_E$ depends on  $\vec M$. This voltage can mimic the so-called ``inverse spin-Hall'' effect.

\section{Approximation of localized spin-flip scattering at the edges ($l_{sf} \ll \ell$)}

The previous section describes the case of negligible spin-flip scattering, i.e. large spin-diffusion length with respect to the width of the Spin-Hall bar. In this section we will explore the {\it opposite limit}, for short spin-diffusion length with respect to the spin-Hall bar : $l_{sf} \ll \ell$. This  situation can be described by a spin-flip scattering that occurs locally at the edges. The approximation of local spin-flip scattering leads to a constant transverse current $J_{y \updownarrow}$, so that simple analytical results can be derived. This approximation is also usual in practice, since the samples are often larger than few tens of microns.

Let us first define $\Delta \mu_0$ such that:
\begin{equation}
\Delta \mu_0 \  =    \  \int_{- \ell}^{+ \ell}  dy \, \partial_{y}
   \mu_ {\uparrow} \  -  \int_{- \ell}^{+ \ell}  dy \, \partial_{y}
   \mu_ {\downarrow}
   \label{SpinPump}
\end{equation}
The subscript $0$ points-out that the potential difference is evaluated between $y = + \ell$ and $y = - \ell$. 
Note that due to the translation invariance along $x$ we have: $ \partial_x \Delta \mu = 0$ and $ \partial_x \mu_\uparrow = \partial_x \mu_\downarrow$.
In the framework of our approximation, the power dissipated by the spin-flip scattering is a constant, given by $ P_\text{sf}  \ = \  v \,\mathcal L\, \Delta \mu_0^2$, where $v = 2 S_{lat} \ell $ is the volume of the device. Inserting the transport equations (\ref{relations-dxmu}) and (\ref{relations-dymu}) into Eq.(\ref{SpinPump}) gives:
\begin{equation}
\Delta \mu_0 \  =\  
  \frac{- \Delta A }{q \eta  (1 + \theta_{so}^2)}
  \end{equation}

where $\Delta A$ is given by the integrals: 
\begin{equation}
\Delta A =    \int_{- \ell}^{+ \ell} dy \, \frac{J_{y \uparrow} + \theta_{so} J_{x \uparrow}}{n_{\uparrow}}
   - \  \int_{- \ell}^{+ \ell} dy \,
   \frac{J_{y \downarrow} - \theta_{so} J_{x \downarrow}}{n_{\downarrow}}.
\label{defA}
\end{equation}

The reduced dissipated power $\tilde{P} = \frac{q \eta (1+\theta_{so}^2)}{S_\text{lat}} \, P$ then reads:
\begin{equation}
    \tilde{P} \  = \  \tilde{P}_J + \tilde{P}_\text{sf} \  = \  \int_{- \ell}^{+ \ell}  \left ( \frac{J_{x \uparrow}^2 + J_{y \uparrow}^2}{n_\uparrow} + \frac{J_{x \downarrow}^2 + J_{y \downarrow}^2}{n_\downarrow} \right ) dy \  + \alpha \ \left (\Delta A \right )^2 
    \label{Power}
\end{equation}
expressed as a function of the control parameter $\alpha$, defined by:
\begin{equation}
\alpha =  \frac{ \mathcal L \, 2 \ell n_0} {\sigma_0 (1 + \theta_{so}^2)} =  \frac{ n_0 \ell}{ l_{sf}^2(1+ \theta_{so}^2)} ,
\label{alpha}
\end{equation}
where we used Eq.(\ref{SF_length}) for the expression of $\alpha$ as a function of the spin-flip scattering length $l_{sf}$. The characteristic length $l_{sf}$ is well-known in the context of the giant magnetoresistance effects. It is usually greater than the Debye-Fermi length $\lambda_D$ and it is range between few nanometers to few micrometers.
 The functional of the dissipated power now reads:
\begin{equation}
{\mathcal{F}}[J_{x \updownarrow},J_{y \updownarrow},n_{\updownarrow}] =   \int_{- \ell}^{\ell} \left [  \left ( \frac{J_{x \uparrow}^2 + J_{y \uparrow}^2}{n_\uparrow} + \frac{J_{x \downarrow}^2 + J_{y \downarrow}^2}{n_\downarrow} \right )  -  \lambda_J \, \left ( J_{x \uparrow} + J_{x \downarrow} \right ) - \lambda_n \, \left ( n_{\uparrow} + n_{\downarrow}  \right ) \right ] dy + \alpha  \left (  \Delta A \right )^2 ,
 \end{equation}
and its minimization leads to the stationarity conditions. The functional derivation as a function of $J_{x \updownarrow}$ gives:
\begin{equation}
  \frac{\delta {\mathcal{F}}}{\delta J_{x \updownarrow}} = 0 \ \Longleftrightarrow \
  \pm 2 \,  \alpha \, \theta_{so} \Delta A + 2  J_{x \updownarrow}  = n_{\updownarrow}  \lambda_{J}
 \label{CondJx}
\end{equation}

The Lagrange coefficient $\lambda_{J}$ is determined using the global conditions Eq.(\ref{ConstrQ}) and Eq.(\ref{ConstrJ}) on currents and charges for the sum of the two channels:
\begin{equation}
 \lambda_{J} =  \frac{2 J_x^0}{n_0 }
    \label{lambda_J}
\end{equation}
 Re-inserting Eq.(\ref{lambda_J}) into Eq.(\ref{CondJx}) gives the longitudinal currents as a function of the spin-dependent density of electric charges:
 \begin{equation}
 J_{x \updownarrow}(y) =   \frac{n_\updownarrow(y)}{n_0 } J_{x}^0  \mp  \alpha \theta_{so} \, \Delta A .
            \label{Jx}
\end{equation}
On the other hand, the minimization as a function of $J_{y \updownarrow}$ gives the current across the spin-Hall bar:
 \begin{equation}
  \frac{\delta {\mathcal{F}}}{\delta J_{y \uparrow}} = 0 \ \Longleftrightarrow \
 J_{y \updownarrow} = \mp \alpha \Delta A 
  \label{Jy0}
\end{equation}
This result shows that the transverse current is a {\it pure spin-current}, in agreement with the results known from the direct resolution of the drift-diffusion equations\cite{Dyakonov,Zhang,Tse,Maekawa,SHE}. Indeed, the charge current vanishes $J_y^Q \equiv J^0_{y \uparrow} + J^0_{y \downarrow} = 0$, and the spin-current reads $J^{Sp}_y \equiv J^0_{y \uparrow} - J^0_{y \downarrow} = 2 J^0_{y \uparrow}$.

Inserting the expression of the transverse current Eq.(\ref{Jy0}) into the expressions of the longitudinal current Eq.(\ref{Jx}) divided by the density $n_{\updownarrow}$, the sum over the two channels then reads:
\begin{equation}
\begin{aligned}
\frac{J_{x \uparrow}}{n_{\uparrow}} + \frac{J_{x \downarrow}}{n_{\downarrow}} =
 \frac{2 J_x^0}{n_0} +  \theta_{so} J_{y}^0 \left  (  \frac{1}{n_{\uparrow}} - \frac{1}{n_{\downarrow}} \right ).
\end{aligned}
\end{equation}
In order to simplify as much as possible the physical interpretation, we continue the derivation at the first order in the accumulation $\delta n_{\updownarrow}/n_0$ (which is a realistic case). This approximation leads also to the first order in $\delta J_{x \updownarrow}/J_x^0$. Equation (\ref{defA}) then becomes $\Delta A \approx \frac{4 \ell}{n_0} \left (J_y^0 + \theta_{so} J_x^0 \right )$ so that the stationarity condition Eq.(\ref{Jy0}) reads:
\begin{equation}
    J_{y \updownarrow}  = \mp \frac{\tilde \alpha}{1+ \tilde \alpha} \theta_{so} J_x^0 
        \label{JySimple}
\end{equation}
where we have introduced the {\it dimensionless parameter }:
\begin{equation}
\tilde \alpha = \frac{4 \ell \alpha}{n_0} = \frac{4 \ell^2}{l_{sf}^2 (1 + \theta_{so}^2)}. 
\label{tilde_alpha}
\end{equation}
The curve $J^{Sp}_y$ (Eq.(\ref{JySimple}) with Eq.(\ref{tilde_alpha})) is plotted in Figure 3a as a function of the ratio $\ell/l_{sf}$. We see that the current $J_y^0$ is composed of the same current as that expected in a Corbino disk of the same material \cite{Benda,Madon}, i.e. $J_{y \updownarrow}^{Corbino} = \mp \theta_{so} J_x^0$, but weighted by the coefficient $\tilde \alpha/(\tilde \alpha + 1)$, where the control parameter $\tilde \alpha$ is given by Eq.(\ref{tilde_alpha}). 

\begin{itemize}

\item
Note that in the case of small spin-flip scattering, i.e. $l_{sf}/\ell > 1$, the control parameter $\tilde \alpha$ goes to zero and $J_{y \updownarrow} \rightarrow 0$, which is indeed the result obtained in the approximation treated in the previous section. 
\item 
In the other limit of strong spin-flip scattering - i.e. small spin-diffusion length $l_{sf}/\ell \ll 1$  - the ratio $\tilde \alpha/(1 + \tilde \alpha)$ is close to $1$, and the spin-currents $J_{y \updownarrow}$ are maximum, like in the Corbino disk. The presence of this transverse pure-spin current with $\tilde \alpha = 1$ is the major feature of the SHE as described in previous theories (i.e. without taking into account the out-of-equilibrium electric screening). The present work describes the transition between the two regimes from large to small values of the parameter $l_{sf}/\ell$.

\end{itemize}

 The transverse charge and spin currents then read:
\begin{equation}
J_{y}^{Q}(y) = 0   \qquad  {\text and } \qquad J_{y}^{Sp}(y)= 2 J^0_{y}  = - 2 \frac{\tilde \alpha}{1+ \tilde \alpha} \theta_{so} \, J_x^0
\label{JySpinCharge}
\end{equation}
where $J_y^0 = J_{y \uparrow}^0 = - J_{y \downarrow}^0$.

\begin{figure} [ht]
   \begin{center}
   \includegraphics[height=7cm]{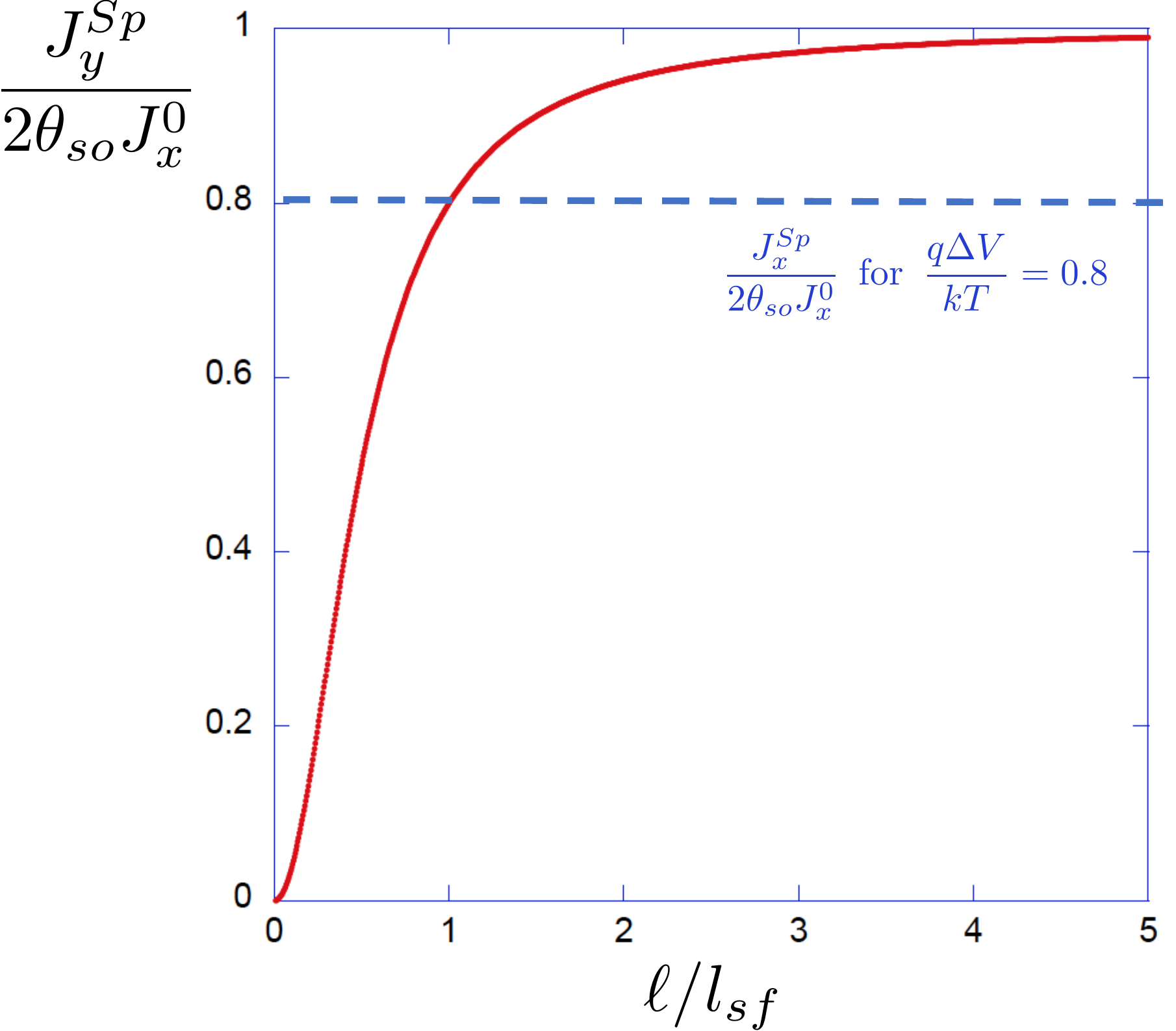}
      \includegraphics[height=7cm]{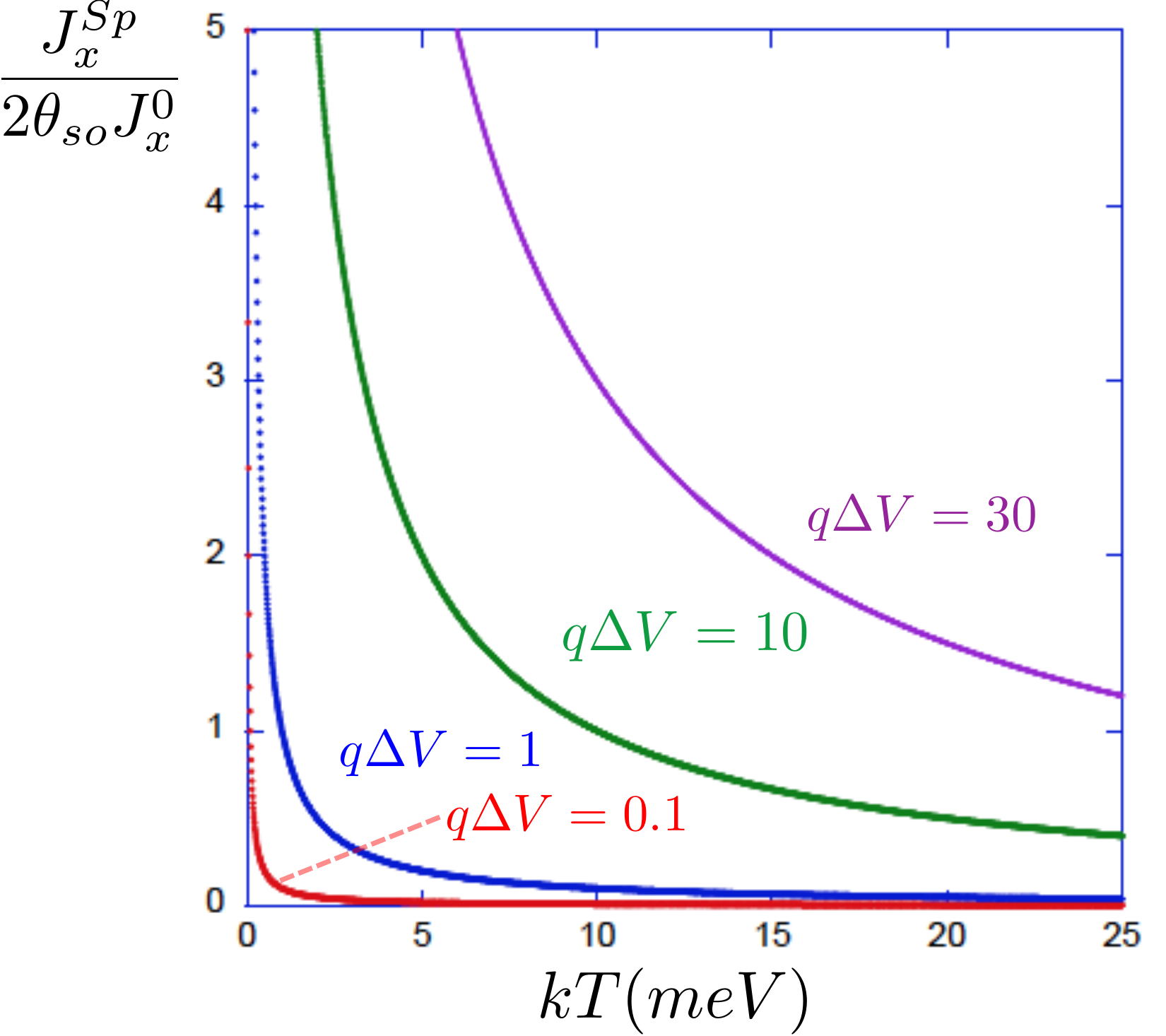}
   \end{center}
   \caption[example] 
   { \label{fig:Fig.3} 
(a) Transverse pure spin current  $J_y^{Sp}$ (Eq.(\ref{JySpinCharge}) in the text)  as a function the ratio $\ell/l_{sf}$ of the width of the Spin-Hall bar over the spin-flip scattering length (Eq.(\ref{JySpinCharge}) and Eq.(\ref{tilde_alpha})). (b)  Longitudinal pure spin current $J_x^{Sp}$ as a function of the temperature (Eq.(\ref{JxCharge})) and (Eq.(\ref{spin-accum})) for different values of the injected voltage $q \Delta V$, expressed in $meV$. The amplitude is calculated for  $y= L = \ell$. The currents are normalized by the product of the injected current $2 J_x^0$ by the spin-Hall angle $\theta_{so}$, at the first order in $\theta_{so}$.}
   \end{figure} 

The longitudinal spin-currents Eq.(\ref{Jx}) with Eq.(\ref{Jy0}) now read:
\begin{equation}
J_{x \updownarrow}(y) =   \left ( \frac{n_\updownarrow(y)}{n_0 }  \mp \frac{\tilde \alpha}{1+ \tilde \alpha}   \theta_{so}^2 \right ) J_{x}^0 .
\label{JxSimple}
\end{equation}
The solution Eq.(\ref{min}) found for $J_{x \updownarrow}$ in the case without spin-flip is now corrected by a small term proportional to $\theta_{so}^2 $. 
The longitudinal charge current and the longitudinal spin current are given by the expressions:
\begin{equation}
J_{x}^{Q}(y)= J_x^0 \, \left (2+ \frac{\delta n^{Q}(y)}{n_0 } \right )
 \qquad  {\text and } \qquad J_{x}^{Sp}(y) = J_x^0 \, \left(  \frac{n^{Sp}(y)}{n_0 } - 2 \frac{ \tilde  \alpha }{1+ \tilde \alpha} \theta_{so}^2  \right )
 \label{JxCharge}
\end{equation}
In conclusion, the stationary state for the spin-Hall bar for $l_{sf} \ll \ell$ and at the first order in $\delta n_{\updownarrow}/n$  is simply defined by equations Eq(\ref{JySimple}) and Eq.(\ref{JxSimple}), where the expression of $n_{\updownarrow}$ is given by the solution of the Poisson law Eq.(\ref{PoissonChem}). At the first order in $\theta_{so}$, the stationary states obtained in the previous section is recovered: there is a solution of continuity between the two opposite approximations used in this study.
The following paragraphs are devoted to the calculation of the densities $n_{\updownarrow}(y)$. \\

Injecting Eq.(\ref{JySimple}) and Eq.(\ref{JxSimple}) into Eq.(\ref{relations-dymu}) in order to obtain $\partial_y\mu$, the Poisson law Eq.(\ref{PoissonChem}) reads now, at the first order in $\delta n_{\updownarrow}/n_0$:
\begin{equation}
 \, \partial_y^2 (\delta n_{\updownarrow}) \mp \frac{\lambda_{SH}}{ \lambda_D^2} \left ( 1 \pm \theta_{so}^2 \right ) \, \partial_y (\delta n_{\updownarrow})  - \frac{\delta n^Q}{\lambda_D^2}   = 0,
 \label{PoissonC}
\end{equation}
where we have define the spin-Hall characteristic length under spin-flip scattering as:
\begin{equation}
\lambda_{SH} \equiv \frac{\epsilon}{q^2 \eta n_0^2} \frac{J_y^0}{1 + \theta_{so}^2} = - \frac{\tilde \alpha  }{1 + \tilde \alpha} \, \lambda_{SH}^0.
\label{Lambda_SH_sf}
\end{equation}
where $\lambda_{SH}^0$ is the spin-Hall length with negligible spin-flip scattering defined in Eq.(\ref{lambdaSH}).
Since the measured Spin-Hall angle verifies $\theta_{so} \ll 1$, Eq.(\ref{PoissonC}) can be approximated at the first order in $\theta_{so}$. Summing and subtracting the equations Eq.(\ref{PoissonC}) for the two channels, we have the coupled equations:
\begin{equation}
 \frac{\partial_y^2 n^{Sp}}{n_0}  - \frac{\lambda_{SH}}{ \lambda_D^2} \, \frac{\partial_y \delta n^Q}{n_0}  = 0,
 \label{EqDiff1}
\end{equation}
\begin{equation}
  \frac{\partial_y^2 \delta n^Q}{n_0} - \frac{\lambda_{SH}}{ \lambda_D^2} \, \frac{ \partial_y n^{Sp}}{n_0}  - \frac{2}{ \lambda_D^2} \, \frac{\delta n^{Q}}{n_0} = 0
  \label{EqDiff2}
\end{equation}

In the case of symmetrical device and electrical environment ($C_E = 0$), the spin-accumulation is a odd function of $y$, and the solutions of equations Eq.(\ref{EqDiff1}) and  Eq.(\ref{EqDiff2}) are:\begin{equation}
\frac{n^{Sp}(y)}{n_0} = a \, y + b \, sh \left (\frac{y}{\lambda_m}  \right )
    \label{Result_Spin}
\end{equation}
and
\begin{equation}
\frac{\delta n^{Q}(y)}{n_0} = c + d \, ch \left (\frac{y}{\lambda_m}  \right )
    \label{Result_Charge}
\end{equation}
where $a$, $b$, $c$ and $d$ are the integration constants defined by the symmetry of the device and the electromagnetic properties of its environment. The fact that the charge accumulation $\delta n^{Q}(y)$ is an even function of $y$ means that the voltage difference $\Delta V = 0$ is zero between the two edges of the device, like in the case without spin-flip (for $C_E = 0$). The {\it typical spin-Hall diffusion length} $\lambda_{m}$ is given by the expression:
\begin{equation}
  \lambda_{m} =  \frac{1}{ \sqrt{ \left (\frac{\tilde \alpha}{1 + \tilde \alpha}  \frac{\lambda_{SH}^0}{\lambda_D^2} \right )^2 +  \frac{2}{\lambda_{D}^2}} }
 \end{equation}

This expression can be simplified due to the small value of the Debye-Fermi length $\lambda_D$. Indeed, the maximal value is of the order of some few microns, so that typically $\lambda_D^{-2} > 10^{10}$ $m^{-2}$. On the other hand, the quantitative evaluation of the ratio $\lambda_{SH}^0/\lambda_D^2 \approx  \theta_{so} \, \left ( \frac{q \Delta V}{kT} \right ) \frac{1}{L}$ has been performed in the previous section below Eq.(\ref{spin-accum}). This quantity depends on the intensity of the voltage $\Delta V$ imposed by the generator, but it is limited for experimental reasons (electromigration), and also due to the limitation of our model (classical or semi-classical diffusive process for which $n^{Sp}/n_0 < 1$). But even if we where able to push the value of  $\lambda_{SH}^0/\lambda_D^2$ to few $10^4$ $m^{-1}$, its square is still negligible with respect to $\lambda_D^{-2}$.  Since we have $\tilde \alpha/(1 + \tilde \alpha) \le 1$ whatever the value of the ratio $l_{sf}/\ell$, we expect $\lambda_{m} \approx \lambda_D/\sqrt 2$ in usual experimental situation. \\

 \begin{figure} [ht!!]
   \begin{center}
   \includegraphics[height=7cm]{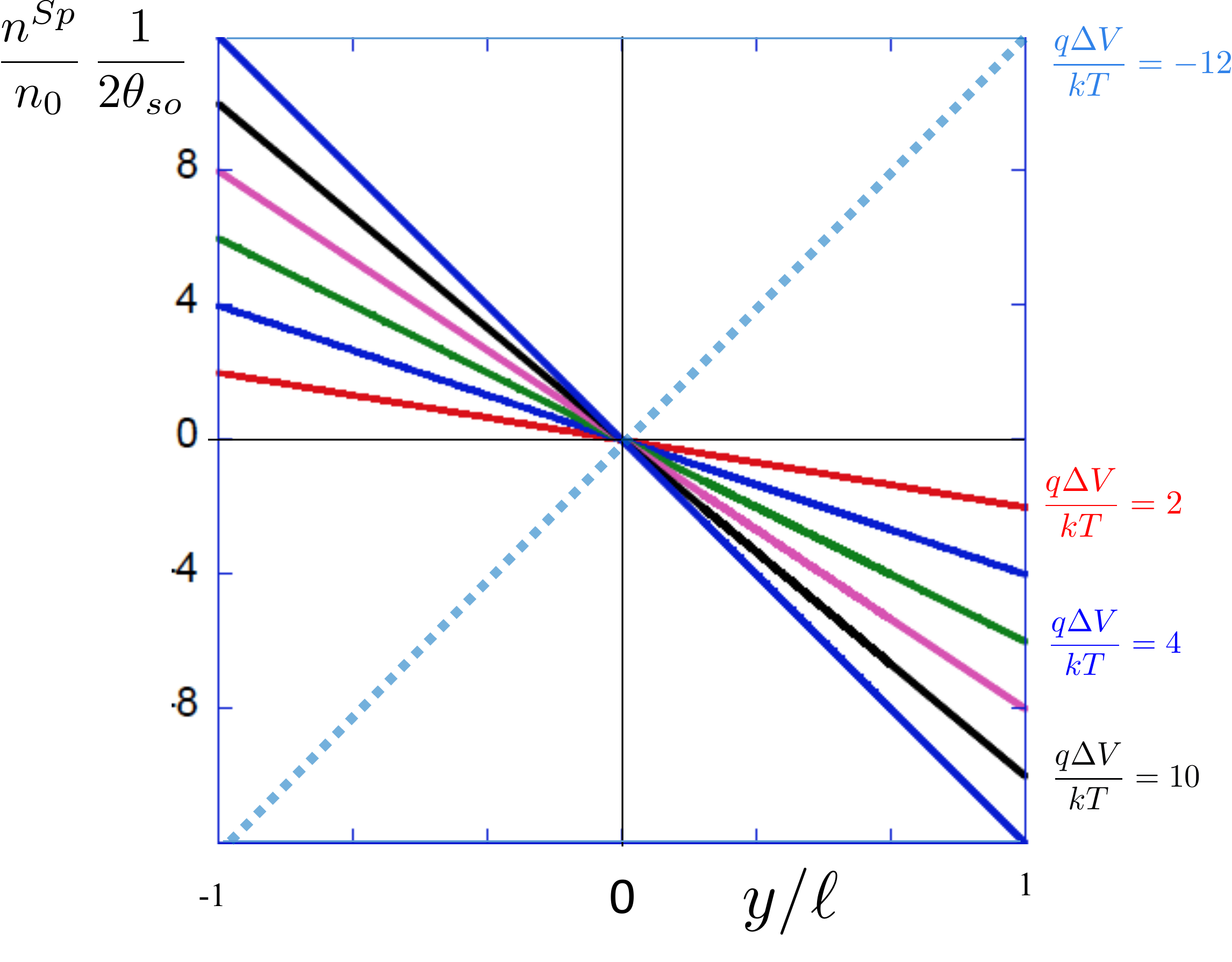}
   \end{center}
   \caption[example] 
   { \label{fig:Fig.3} 
Spin accumulation $n^{Sp}$ (Eq.(\ref{spin-accum}) in the text) as a function the ratio $y/\ell$ for different values of the ratio $q \Delta V/k T$, from $q \Delta V/k T= 2$ to $q \Delta V/k T = 12$. The spin-accumulation is normalized by the product of the density of carriers at equilibrium $2n_0$ by the spin-Hall angle $\theta_{so}$. The curve for $q \Delta V/k T= -12$ shows that reversing the electric polarity reverses the sign of the spin-accumulation.}
   \end{figure}

Taking into account the symmetry of the device and its environment ($C_E = 0$), the global condition Eq.(\ref{ConstrQ}), and the result of section III for the limit $\ell/l_{sf} \ll 1$, we have:
\begin{equation}
a = - 4 \frac{\lambda_m^2}{\lambda_D^2} \, \left ( \frac{ \lambda_{SH}^0}{ \lambda_D^2} \right ) \approx -2 \frac{ \lambda_{SH}^0}{ \lambda_D^2},
\label{cst_a}
\end{equation}
so that the first term appearing in the right-hand side of Eq.(\ref{Result_Spin}) corresponds the spin-accumulation without spin-flip obtained for Eq.(\ref{spin-accum}) in the previous section. The second term is given by:
\begin{equation}
b = - 2 \frac{\ell  \lambda_m^2 }{sh\left (\frac{\ell}{\lambda_m} \right )} \, \left ( \frac{ \lambda_{SH}^0}{ \lambda_D^2} \right )^3 \, \left ( \frac{\tilde \alpha}{1 + \tilde \alpha} \right )^2 .
\label{cst_b}
\end{equation}
The non-linear correction $b \, sh(y/\lambda_m)$ appearing in Eq.(\ref{Result_Spin}) has a maximum value of the order of  $\lambda_D^2 \, a^2 $, much smaller than the linear part (in usual experimental situations), which is of the order of $\ell \, a $. Equation (\ref{Result_Spin}) thus reduced to
$\frac{n^{Sp}(y)}{n_0} \approx a \, y $, which is the result obtained in the previous section for the case without spin-flip, as shown in Figure 3b  and Figure 4.
On the other hand, the constant for the charge accumulation Eq.(\ref{Result_Charge}) is:
\begin{equation}
c = - 2 \frac{\tilde \alpha}{ 1 + \tilde \alpha} \, \lambda_m^2 \, \left ( \frac{ \lambda_{SH}^0}{ \lambda_D^2} \right )^2,
\label{cst_c}
\end{equation}
which maximum value of the order of $\lambda_D^2 a^2$, i.e. small, and
\begin{equation}
d = 2 \frac{  \lambda_m \ell}{sh\left (\frac{\ell}{\lambda_m} \right )} \, \left ( \frac{ \lambda_{SH}^0}{ \lambda_D^2} \right )^2 \, \left ( \frac{\tilde \alpha}{1 + \tilde \alpha} \right ), 
\label{cst_b}
\end{equation}
so that the maximum value of the second term $d \, ch(y/\lambda_m)$ in the right-hand side of Eq.(\ref{Result_Charge}) is also of the order $\lambda_D^2 a^2$. \\

Accordingly, it appears that {\bf  the role of the spin-flip scattering is negligible for both spin accumulation and charge accumulation}. The unique sizable spin-flip scattering effect in the SHE is thus the generation of the transverse spin-current $J_y^{Sp}$ (already well-known in the usual drift-diffusion description of the SHE). The result about the linear dependence on $y$ confirms the observations performed by Bottegoni et al. \cite{Bottegoni} on devices for which the spin-diffusion length $l_{sf}$ is large in absolute value, but still smaller or of the same order than the width $\ell$ of the spin-Hall bar. The measurements show that the spin-accumulation is linear in $y$, linear in $\Delta V$, and inversely proportional to the temperature.
\\

\section{Conclusion}

The stationary state of the spin-Hall bar has been studied in the framework of a variational approach that includes non-equilibrium screening effects. The minimization of the Joule power dissipated in the system is performed with taking into account the spin-flip relaxation and the two global constrains (global galvanostatic conditions and global electroneutrality).
The calculation is performed within the two limiting cases that are the negligible spin-flip scattering limit $l_{sf}/\ell \gg 1$ and the strong spin-flip scattering limit $l_{sf}/\ell \ll 1$ (where $l_{sf}$ is the spin-flip scattering length and $\ell$ is the width of the Spin-Hall bar). In both cases, the profile of the spin-accumulation $n^{Sp}(y)$ and the spin-currents $\vec J_{\updownarrow}(y)$ can be described analytically. The two limits coincide, and simple expressions are given at the first order in the spin-accumulation.

In the approximation of negligible spin-flip scattering, the main result is the absence of transverse currents $J_{y \updownarrow} = 0$ with a longitudinal spin current $J_{x}^{Sp}(y)$ proportional to the spin accumulation $n^{Sp}(y)$, while the charge current $J_x^Q$ is zero (which is the usual definition of a ``pure spin current''). 
 The spin-accumulation $n^{Sp}(y)$ is shown to be linear in $y$ (across the device), linear in the electric field imposed by the generator along the $x$ axis (i.e. linear to the injected current $J_x^0$), and inversely proportional to the temperature for non-degenerate conductors. Note that the temperature dependence that is given by the ratio $\theta_{so}/kT$ is the out-of-equilibrium reminiscence of the Curie law that describes a paramagnetic system, for which the external magnetic field has been replaced by a weak internal spin-orbit field.  \\

In the case of strong spin-flip scattering, the main difference with the case of negligible spin-flip is the presence of a transverse pure spin-currents $J_{y}^{Sp}$, which is flowing across the sample (as predicted by the drift-diffusion theory). This current is weighted by the factor $\tilde \alpha/(1+ \tilde \alpha)$ (Eq.(\ref{JySimple})), where $\tilde \alpha \approx (2 \ell /( l_{sf})^2 $ (Eq.(\ref{alpha})). This multiplying factor describes the transition from a strong spin-flip scattering regime to a weak spin-flip scattering regime, for which the transverse spin-current vanishes. 
 
The most surprising result of this study is that the spin-flip scattering does not change significantly the stationary state of the Spin-Hall effect (for standard experimental situations), except by the presence of the a transverse pure-spin-current $J_y^{Sp}$. Indeed, the longitudinal spin current $J_{x}^{Sp}(y)$ is approximatively the same as for the case without spin-flip scattering for two reasons. The first reason is that the supplementary term due to spin-flip scattering is at the second order in the spin-Hall angle $\theta_{so}^2$, and this term is negligible in general.  The second reason is that the spin-accumulation  $n^{Sp} (y)$ is also approximatively unchanged whatever the value of the ratio $\ell/l_{sf}$, due to the small value of the Debye-Fermi length scale $\lambda_D$. 
 
Furthermore, it is shown that if the Spin-Hall bar is asymmetric, a spin-dependent voltage is generated between the two edges. This voltage could then be measured if a layer (either oxide, semiconductor, or metal) is deposited on one side of the device. \\

In conclusion, in all cases, the stationary state is defined by a spin-accumulation $n^{Sp}(y)$ (and the inhomogeneous part of the pure-spin-current $J_x^{Sp} \propto n^{Sp}(y)$) which is linearly distributed across the spin-Hall-bar (i.e. along the $y$ direction), linear in the applied electric potential $\Delta V$, and proportional to the inverse of the temperature for non-degenerate conductors. These surprising characteristics are observed in the measurements performed by Bottegoni et al \cite{Bottegoni} in a series of measurements, for which $\ell/ l_{sf}$ is slightly greater than one


\section{Acknowledgement} 
We thank Felix Faisant for his important contribution to the initial development of this work, Pierre-Michel d\'ejardin for the stochastic description, and Serge Boiziau for his help and for valuable discussions.\\


\begin{thebibliography}{50}


\bibitem{Awschalom} Y. K. Kato, R. C. Myers, A. C. Gossard, D. D. Awschalom, {\it Observation of the spin Hall effect in semiconductors}, Science {\bf 306} 1910 (2004). 
\bibitem{Jungwirth} J. Wunderlich; B. Kaestner; J. Sinova; T. Jungwirth, {\it Experimental Observation of the Spin-Hall Effect in a Two-Dimensional Spin-Orbit Coupled Semiconductor System}, Phys. Rev. Lett. {\bf 94}, 047204  (2005).
\bibitem{Valenzuela} S. O. Valenzuela and M. Tinkham {\it Direct electronic measurement of the spin-Hall effect}, Nature {\bf 442}, 176 (2006).
\bibitem{Otani} T. Kimura; Y. Otani; T. Sato; S. Takahashi; S. Maekawa {\it Room-Temperature Reversible Spin Hall Effect}, Phys. Rev. Lett. {\bf 98}, 156601 (2007).
\bibitem{Gambardella} C. Stamm, C. Murer, M. Berritta, J. Feng, M. Gaburac, P. M. Oppeneer, and P Gambardella, {\it Magneto-optical detection  of the Spin Hall effect in Pt and W thin films}, Phys. Rev. Lett. {\bf 119}, 087203 (2017).
\bibitem{Bottegoni} F. Bottegoni, C. Zucchetti, S. Dal Conte, J. Frigerio, E. Carpene, C. Vergnaud, M. Jamet, G. Isella, F. Ciccacci, G. Cerullo, and M. Finazzi, {\it Spin-Hall Voltage over a Large Length Scale in bulk Germanium}, Phys. Rev. Lett. {\bf 118}, 167402 (2017).
\bibitem{Dyakonov} M. I. Dyakonov, and  V. I. Perel, {\it Possibility of orienting electron spins with current} ZhETF Pis. Red. {\bf 13}, no 11, 657-660 (1971) 
\bibitem{Dyakonov2} M. I. Dyakonov, {\it spin Physics in Semiconductors}, Springer Series in Solid-States Sciences 2008.
\bibitem{Hirsch} J. E. Hirsch {\it Spin Hall effect} Phys. Rev. Lett. {\bf 83}, 1834 (1999). 
\bibitem{Zhang} Sh. Zhang, {\it Spin Hall effect in the presence of spin diffusion}, Phys. Rev. Lett. {\bf 85}, 393 (2000).
\bibitem{Tse}  W.-K. Tse, J. Fabian I. $\check{Z}$uti\'c, and S. Das Sarma, {\it Spin accumulation in the extrinsic spin Hall effect}, Phys. Rev. B {\bf 72}, 241303(R) (2005)
\bibitem{Maekawa} S. Takahashi and S. Maekawa {\it Spin current, spin accumulation and spin Hall effect}, Sci. Technol. Adv. Mater. {\bf 9} (2008) 014105. 
\bibitem{Hoffmann} A. Hoffmann, {\it Spin Hall effects in metals}, IEEE Trans. Mag. {\bf 49} (2013) 5172.
\bibitem{Saslow} W. M. Saslow, {\it Spin Hall effect and irreversible thermodynamics: Center-to-edge transverse current-induced voltage}, Phys. Rev .B {\bf 91}, 014401 (2015).
\bibitem{SHE} J. Sinova, S. O. Valenzuela, J. Wunderlich, C. H. Back, T. Jungwirth  {\it Spin Hall effects}, Rev. Mod. Phys. {\bf 87}, 1213 (2015).
\bibitem{Sinova} J. Sinova, T. Jungwirth {\it Surprises from the spin-Hall effect}, Physics Today {\bf 70}, 7, 38 (2017).
\bibitem{boundary} M. J. Moelter, J. Evans, G. Elliott, and M. Jackson, {\it Electric potential in the classical Hall effect: An unusual boundary-value problem}, Am. J. Phys. {\bf 66}, 668 (1998)

\bibitem{Trefenthen} L.N. Trefenthen, R.J. Williams, {\it Conformal mapping solution of Laplace's equation on a polygon with oblique derivative boundary conditions},  J. Comput. Appl. Math. {\bf 14}, 227-249 (1986).

\bibitem{Heremans} D.R. Baker, J.P. Heremans, {\it Linear geometrical magnetoresistance effect : Influence of geometry and material composition}, Phys. Rev. B {\bf 59}, 13927 (1999).

\bibitem{Geometry} Guo Zhang, Jinyu Zhang, Zhan Liu, Peng Wu, Huaqiang Wu, He Qian, Yan Wang, Zhiyong Zhang, and Zhiping Yu {\it Geometry Optimization of Planar Hall Devices Under Voltage Biasing}, IEEE Transactions on electron devices, {\bf 61}, 4216, (2014).

\bibitem{Connection} Oliver Paul and Martin Cornil, {\it Explicit connection between sample geometry and Hall response}, Appl. Phys. Lett. {\bf 95}, 232112 (2009).

\bibitem{Perrott} Tobias Kramers, Viktor Krueckl, Eric. J. Heller, and Robert E. Parrott {\it Self-consistent calculation of electric potentials in Hall devices}, Phys. Rev. B {\bf 81}, 205306 (2010).

\bibitem{Nanowire} C. Fernandes, H. E. Ruda, and A. Shik, {\it Hall effect in nanowires}, J. Appl. Phys. {\bf 115}, 234304 (2014).

\bibitem{Calcul} D. Homentcovschi and R. Bercia, {\it Analytical solution for the electric field in Hall plates},  Z. Angew. Math. Phys. 69:97 (2018).

\bibitem{Solin} S. A. Solin, Tieneke Thio, D. R. Hines, J. J. Heremans, {\it Enhanced Room-Temperature Geometric Magnetoresistance in Inhomogeneous Narrow-Gap Semiconductors}, Science {\bf 289}, 1530 (2000).

\bibitem{SolinJAP} Lisa M. Pugsley, L. R. Ram-Mohan, and S.A. Solin, {\it Extraordinary magnetoresistance in two and three dimensions: Geometrical optimization}, J. Appl. Phys. {\bf 113}, 064505 (2013).

\bibitem{EPL1} J.-E. Wegrowe, R. V. Benda, and J. M. Rub\`i., {\it Conditions for the generation of spin current in spin-Hall devices}, Europhys. Lett {\bf 18} 67005 (2017).

\bibitem{EPL2} J.-E. Wegrowe, P.-M. Dejardin, {\it  Variational approach to the stationary spin-Hall effect},  Europhys. Lett  {\bf 124}, 17003 (2018).

\bibitem{Benda} R. Benda, E. Olive, M. J. Rub\`i and J.-E. Wegrowe {\it Towards Joule heating optimization in Hall devices}, Phys. Rev. B  {\bf 98}, 085417 (2018).

\bibitem{JAP1}  M. Creff, F. Faisant, M. Rub\`i, J.-E. Wegrowe {\it Surface current in Hall devices}, J. Appl. Phys. {\bf 128},  054501 (2020). 
https://doi.org/10.1063/5.0013182.
 
 \bibitem{JAP2}   P.-M. D\'ejardin and J-E. Wegrowe {\it Stochastic description of the stationary Hall effect}, J. Appl. Phys. {\bf 128},  184504 (2020).
\bibitem{JAP3} F. Faisant, M. Creff, J.-E. Wegrowe {\it The physical properties of the Hall current}, J. Appl. Phys. {\bf 129},  144501 (2021).
\bibitem{Onsager_Diss} L. Onsager and S. Machlup, {\it Fluctuations and irreversible processes}, Phy. Rev. {\bf 91} 1505 (1953).

\bibitem{Bruers} S. Bruers, Ch. Maes, K. Netocn\'y, {\it On the validity of entropy production principles for linear electrical circuits}, J. Stat. Phys. {\bf 129}, 725 (2007).

\bibitem{MinDiss} L. Bertini, A. De Sole, D. Gabrielli, G. Jona-Lasinio, and C. Landim {\it Minimum Dissipation Principle in Stationary Non-Equilibrium Sates}, J. Stat. Phys. {\bf 116}, 831 (2004).

\bibitem{Hall} E. H. Hall {\it On a new action of the magnet on electric currents}, Am. J. Math. \textbf {2}, 287 (1879).


\bibitem{Johnson} M. Johnson and R. H. Silsbee, {\it "Interfacial charge-spin coupling: injection and detection of spin magnetization in  metal"}, Phys. Rev. Lett. {\bf 55}, 1790 (1985) 
\bibitem{Wyder} P.C. van son, H. van Kempen, and P. Wyder, {\it Boundary resistance of the ferromagnetic-nonferromagnetic metal interface}  Phys. Rev. Lett. {\bf 58}, 2271 (1987)
\bibitem{Fert} T. Valet and A. Fert, {\it Theory of the perpendicular magnetoresistance in magnetic mutilayers}, Phys. Rev. B {\bf 48}, (1993)
\bibitem{PRB2000} J.-E. Wegrowe, {\it Thermokinetic approach of the generalized Landau-Lifshitz-Gilbert equation with spin-polarized current}, Phys. Rev. B {\bf 62}, (2000), 1067. 
\bibitem{Schmidt} G. Schmidt, D. Ferrand, L. W. Molenkamp, A.T. Filip and B.J. van Wees, {\it Fundamental obstacle for electrical spin injection from a ferromagnetic metal into a diffusive semiconductor}, Phys. Rev. B {\bf  62}, R4790 (2000).
\bibitem{Jedema} F. J. Jedema, M. S. Nijboer, A. T. Filip, and B. J. van Wees, {\it Spin injection and spin accumulation in all-metal mesoscopic spin-valves}, Phys. Rev . B {\bf 67}, 085319 (2003).

\bibitem{JPhys17} J.-E. Wegrowe {\it Twofold stationary states in the classical spin-Hall effect} J. Phys.: Condens. Matter {\bf 29} 485801 (2017).

\bibitem{RqueT} In the case of degenerate semiconductors and metals, the expression is valid at the first order in $\delta n/n_0$.

\bibitem{Rubi} D. Reguera, J. M. G. Vilar, and J. M. Rub\`i, {\it The mesoscopic Dynamics of Thermodynamic Systems}, J. Phys. Chem. B {\bf 109} (2005).

\bibitem{Onsager} L. Onsager {\it Reciprocal relations in irreversible processes II}, Phys. Rev. {\bf 38}, 2265 (1931)

\bibitem{Beach} A. Churikova, D. Bono, B. Neltner, A. Wittmann, L. Scipioni, A. Shepard, T. Newhouse-Illige, J. Greer, and G.S.D. Beach {\it Non-magnetic origin of SH Magnetoresistance-like signals in Pt films and epitaxial NiO/Pt layers}, J. Appl. Phys. {\bf  116} 022420 (2020).


\bibitem{DiVentra} Yuriy V Pershin and Massimiliano Di Ventra, {\it A voltage probe of the spin Hall effect}, J. Phys.:Condens. Matter {\bf 20} (2008) 025204,
 doi:10.1088/0953-8984/20/02/025204


\bibitem{Zucchetti17} C. Zucchetti, F. Bottegoni, C. Vergnaud, F. Ciccacci, G. Isella, L. Ghirardini, M. Celebrano, F. Rortais, A. Ferrari, A. Marty, M. Finazzi, and M. Jamet
{\it Imaging spin diffusion in germanium at room temperature}, Phys. Rev. B {\bf 96}, 014403 (2017).


\bibitem{Madon}  B. Madon, M. Hehn, F. Montaigne, D. Lacour, and J.-E. Wegrowe.  {\it Corbino magnetoresistance in ferromagnetic layers : Two representative examples $Ni_{81}Fe_{19}$ and $Co_{83}Gd_{17}$ } Phys. Rev B (R) {\bf  98} 220405(R) (2018).


\end{thebibliography}
\end{document}